\documentclass[10pt,twocolumn]{article}
\usepackage{graphicx}
\usepackage{setspace}
\usepackage{amsmath}
\usepackage{amssymb}
\usepackage{graphicx}
\usepackage{stfloats}
\usepackage{placeins}
\usepackage{fullpage}
\usepackage{abstract}
\usepackage{authblk}
\usepackage[superscript,biblabel]{cite}
\newcommand\blfootnote[1]{%
  \begingroup
  \renewcommand\thefootnote{}\footnote{#1}%
  \addtocounter{footnote}{-1}%
  \endgroup
  
    \renewcommand{\topfraction}{0.9}	
    \renewcommand{\bottomfraction}{0.8}	
    \setcounter{topnumber}{2}
    \setcounter{bottomnumber}{2}
    \setcounter{totalnumber}{4}     
    \setcounter{dbltopnumber}{2}    
    \renewcommand{\dbltopfraction}{0.9}	
    \renewcommand{\textfraction}{0.07}	
    \renewcommand{\floatpagefraction}{0.7}	
    \renewcommand{\dblfloatpagefraction}{0.7}	
}

\begin{document}

\title{Observation of rogue events in non-Markovian light}%
\author{Hadas Frostig$^{1,\dag,*}$, Itamar Vidal$^{1,2,\dag}$,Robert Fischer$^3$, Hanan Herzig Sheinfux$^4$, and Yaron Silberberg$^{1,5}$}

\affil{\footnotesize $^1$Department of Physics of complex Systems, Weizmann Institute of Science, Rehovot 76100, Israel \\$^2$ 2LOQEF,  57039-739 Maceió, Alagoas, Brazil \\$^3$CPGEI, Federal University of Technology - Paran\'a, 80230-901 Curitiba, PR, Brazil \\$^4$ Physics Department, Technion, Israel Institute of Technology, Haifa 32000, Israel \\$^5$ Deceased}

\twocolumn[
  \maketitle
\begin{onecolabstract}
The efforts to understand the physics of rogue waves have motivated the study of mechanisms that produce rare, extreme events, often through analogous optical setups. As many studies have reported nonlinear generation mechanisms, recent work has explored whether optical rogue events can be produced in linear systems. Here we report the observation of linear rogue events with tunable height, generated from light imprinted with a non-Markovian wavefront. Moreover, if the non-Markovian wavefront is allowed to propagate through a nonlinear medium, extraordinarily long-tailed intensity distributions are produced, which do not conform to the statistics previously observed in optical rogue wave experiments.
\vspace{1cm}
\end{onecolabstract}
]
\section{Introduction}
\blfootnote{\dag These authors contributed equally to this work.} 
\blfootnote{*Corresponding author: hfrostig@bu.edu.}
\vspace{-1cm}
Over the past centuries, rare incidents of unexpected, extraordinarily large oceanic waves have been reported, together with their devastating consequences \cite{Dysthe2008}. The effort to understand the mechanisms driving the generation of such events, referred to as rogue waves, has motivated the study of systems with similar behavior, particularly in optics. In a pioneering experiment, Solli \emph{et al.} observed the generation of optical rogue waves in nonlinear media with modulation instability \cite{Solli2007}. This demonstration opened the door to much research on the generation of optical rogue waves in nonlinear media. Optical rogue waves generated with modulation instability were modeled \cite{Baronio2012,Toenger2015} and observed in several systems, such as optical fibers \cite{Kibler2010}, supercontinuum generation \cite{Mussot2009,Erkintalo2009}, Raman fiber amplifiers \cite{Hammani2008}, photorefractive crystals \cite{Pierangeli2015}, nonlinear optical cavities \cite{Montina2009}, and femtosecond filamentation \cite{Dubietis2011,Birkholz2013}. Furthermore, the interaction of nonlinearity and turbulence \cite{Agafontsev2015,Horowitz1997} was studied with random light sources \cite{suret2016}, Kerr resonators \cite{Coulibaly2019}, and ring-cavity lasers \cite{Lecaplain2014}. Optical rogue waves have been demonstrated also in dissipative nonlinear systems such as femtosecond lasers \cite{Lecaplain2012,SotoCrespo2011} and fiber lasers \cite{Liu2016}.

In light of these results, the question arose as to what role does nonlinearity play in generating rare, extreme events and whether it is a fundamental requirement for their existence \cite{onorato2013,Dudley2014,Dudley2019}. Deviations from Rayleigh statistics, a criteria that the optical community often uses for the existence of rogue waves, can be observed when the conditions of Rayleigh scattering are violated \cite{Weng1992}. Intensity distributions obeying Rayleigh statistics are generated when a homogeneous wave is Rayleigh scattered by a large number of random scatterers, creating a uniform distribution of phases in the interval $[-\pi,\pi]$ in the scattered wave, and the scattered wave is observed at a distance from the scattering layer (fully-developed speckles) \cite{goodman2000}. The distribution of intensities in the scattered wave is a decaying exponential, conventionally plotted as a linear distribution in semilog scale. By devising systems that defy the conditions of Rayleigh scattering, long-tailed intensity distributions have been theoretically predicted \cite{Nieuwenhuizen1995,Metzger2014} and observed in various systems without the use of nonlinearity, including microwave resonators \cite{Hohmann2010}, nanowire mats \cite{Strudley2013}, partially developed speckle fields \cite{Mathis2015}, inhomogeneously illuminated scatterers \cite{Arecchi2011} and speckle fields with strong phase modulations \cite{Apostol2003,Safari2017}. Particularly, wavefront shaping provides a means of generating non-uniform phase distributions \cite{Liu2015} in a controllable manner, and has been used to generate long-tailed distributions by numerically retrieving the phase mask necessary to create them \cite{Bromberg2014}. 

Here we study an optical wave model that uses uniform illumination of a large number of random scatterers, yet does not lead to a Rayleigh distribution in the far-field. The generation mechanism we study is non-Markovian \cite{Fischer2015,Eichelkraut2015}, inspired by the non-Markovian behavior of dynamical turbulent systems in the oceanic environment, such as sea surface winds \cite{delsole2000,sura2003,thompson2014}. Non-Markovian distributions are distributions with long-range correlations and hence some degree of memory. We show that a non-Markovian light source creates rare, extreme events even in the linear regime, yet their intensities are drastically enhanced by nonlinearity in the medium through which they propagate. Our system allows for tunability of the height of the rogue events, which can be used to study their behavior in different regimes, including a regime with statistics resembling generalized extreme value (GEV) statistics.

\section{Principle}
\begin{figure}[htbp]
\centering
\includegraphics[width=\linewidth]{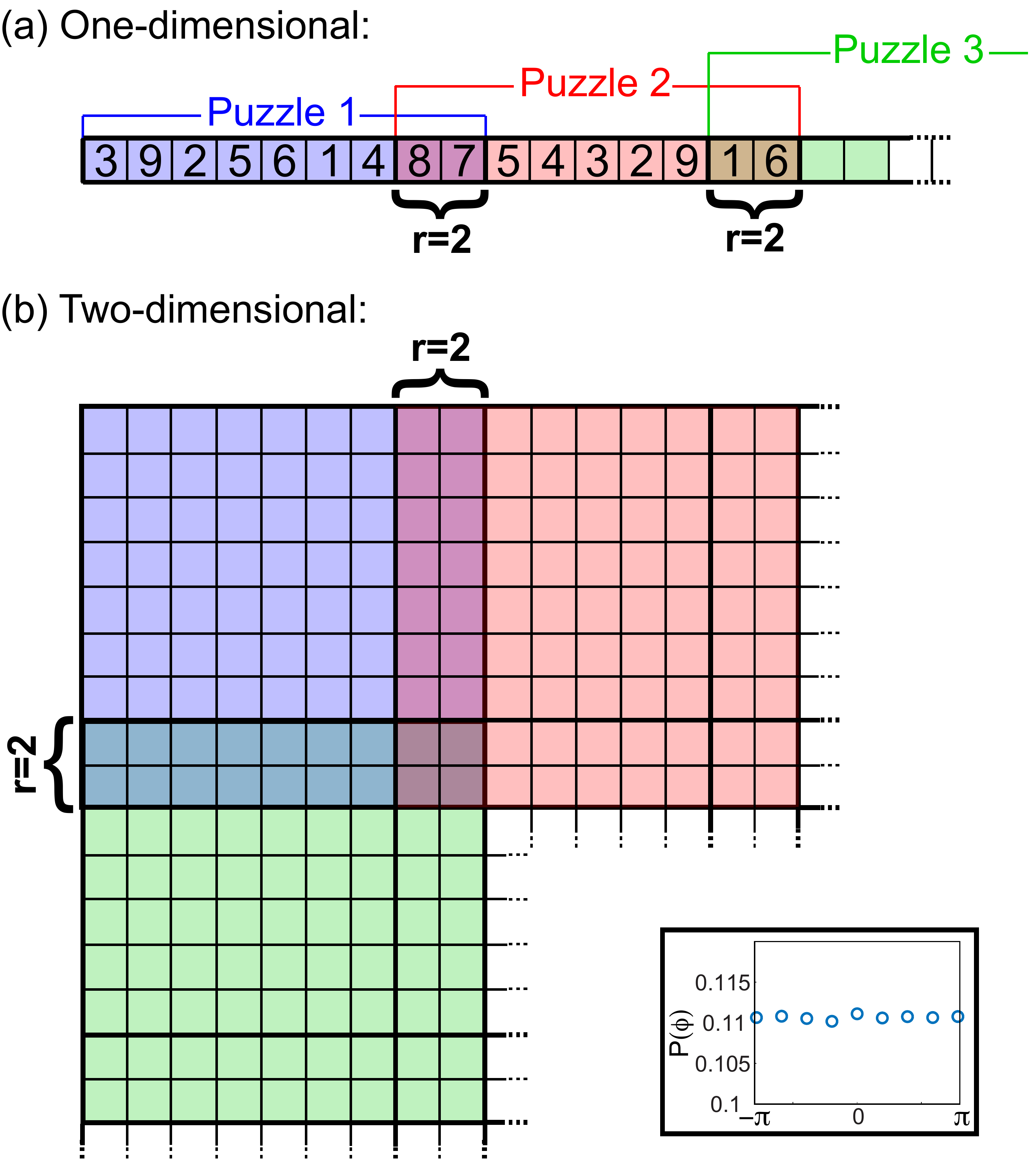}
\caption{\footnotesize\textbf{Generating non-Markovian phase masks.} (\textbf{a}) A one-dimensional non-Markovian phase mask is generated by solving overlapping one-dimensional Sudoku puzzles of length 9. In each such puzzle the number 1-9 should appear only once. (\textbf{b}) Two-dimensional non-Markovian phase pattern are similarly generated by solving overlapping two-dimensional Sudoku puzzles of length 9. In each such puzzle the number 1-9 should appear only once in each row and each column. The cases shown in a and b are for an overlap of r=2. Inset: the probability distribution of the phases in the r=7 non-Markovian phase mask. The phase values are uniformly distributed in the range $[-\pi,\pi]$ despite the non-Markovian properties.}
\label{PhasePatt}
\end{figure}

To generate non-Markovian light, we imprinted a coherent wavefront with a two-dimensional spatial phase mask containing random phases with a tunable amount of long-range order. A number set is said to be non-Markovian when the value at a given position within the set is statistically dependent on numbers at other positions, in addition to its nearest neighbors on both sides. Such a set would result, for example, from drawing distinct balls from a sac without replacing them, so that the outcome of each draw depends on all previous draws.
In order to produce a two-dimensional phase pattern that would be non-Markovian on the one hand, and uniformly distributed within the range $[-\pi,\pi]$ on the other hand, we followed the procedure described by Fischer \textit{et al.} \cite{Fischer2015} and illustrated in Fig.~\ref{PhasePatt}. In short, the phase values are chosen by solving overlapping Sudoku puzzles. Each puzzle is a 9x9 square in which the numbers 1-9 must appear only once in each row and each column. After the values of a single square have been assigned, the next square is defined such that some of its columns (or rows) overlap with the previous square. We denote the number of overlapping columns (or rows) by r. The larger r is, the more memory there is in the generated pattern and consequently the larger the degree of non-Markovianity. Note that r = 9 - s, where s is the shift parameter defined by Fischer \textit{et al.}. This procedure is repeated until the entire matrix is filled, resulting in a random correlated number set \cite{keen2015,Schorr2003}. This set of numbers, N, is then translated into phase values, $\phi$, according to $\phi = 2\pi \frac{N}{9}$. The inset of Fig.~\ref{PhasePatt} shows the probability distribution of the phases of the r=7 mask, which has the largest degree of long-range correlation. We see that adding long-range order to the random phase masks indeed does not modify the uniform distribution of their phases in the range $[-\pi,\pi]$. 

Our experimental setup is outlined in Fig.~\ref{Setup} below. The generated non-Markovian phase mask is imprinted onto light from a continuous-wave laser with a two-dimensional spatial light modulator. The shaped wavefront undergoes Fourier transform by a lens and then propagates through a photorefractive crystal (SBN:60, see methods). Since the nonlinear refractive index of the crystal depends on the strength and sign of the electric field applied to it, the crystal is used as either a linear, positive nonlinear, or negative nonlinear medium. The output facet of the crystal is imaged onto a CCD camera and the recorded intensity distribution is analyzed. The experiment is repeated for several overlap amounts, namely r=0,4,7. 

\begin{figure}[htbp]
\centering
\includegraphics[width=\linewidth]{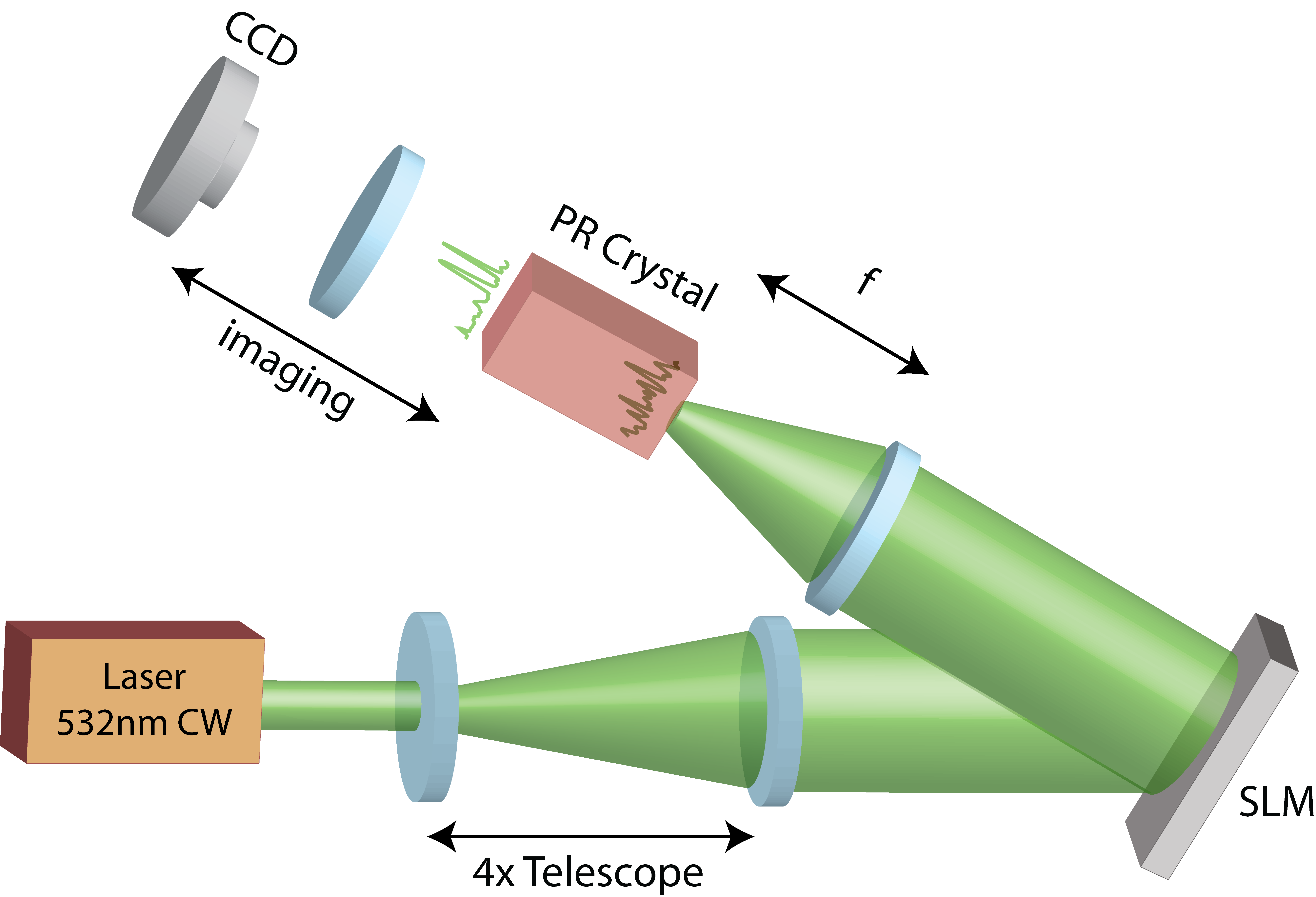}
\caption{\footnotesize\textbf{Generating rogue events with a non-Markovian light source.} A 532nm laser beam is expanded using a telescope to illuminate a two-dimensional SLM. The SLM is used to imprint a non-Markovian spatial phase onto the beam. The beam is then focused into a photorefractive crystal. The output facet of the crystal is imaged onto a CCD camera.}
\label{Setup}
\end{figure}

\section{Results}
\subsection{Experimental results}
\begin{figure*}[htbp]
\centering
\includegraphics[width=\textwidth]{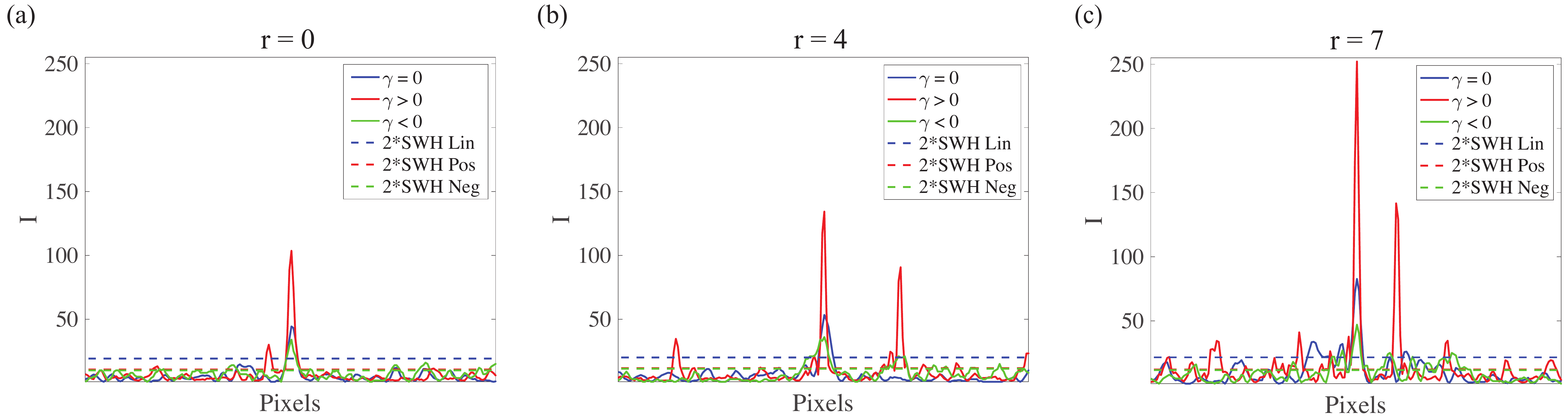}
\caption{\footnotesize\textbf{Experimental results: One-dimensional cuts through the recorded two-dimensional intensity patterns.} (a) A 1D cut through the intensity pattern recorded by the CCD for the r=0 phase pattern, after propagation through a linear (blue), positive nonlinear (red) and negative nonlinear (green) medium. (b) Similarly, for the r=4 phase pattern. (c) Similarly, for the r=7 phase pattern. Dashed lines at the colors corresponding to each distribution are plotted in all figures at two times the significant wave height, denoting the threshold value for rogue waves. The combination of non-Markovianity and nonlinearity produces extremely large events, whose height can be tuned with the overlap parameter, r.}
\label{1D}
\end{figure*}

For initial evaluation of the experimental results, we plotted 1D slices of the recorded 2D output intensity patterns for different conditions. Fig.~\ref{1D} compares 1D slices taken at the location of the maximum intensity of the output patterns for the cases of propagation through a linear (blue), positive nonlinear (red), and negative nonlinear (green) media, for various overlap values. The r=0 case (Fig.~\ref{1D}a) represents the non-Markovian phase mask with the least amount of order, since the phase of a given pixel is statistically dependent only on the other pixels within its own 9x9 square. In contrast, the r=7 case (Fig.~\ref{1D}c) represents the non-Markovian phase mask with the most order, since the large overlap between the squares places the largest number of constraints on the value of a given pixel possible, while maintaining random assignment. An established criterion for rogue waves is whether their height exceeds two times the significant wave height (SWH), defined as the mean amplitude of the highest third of the wave events \cite{Dudley2014,akhmediev2016}. Dashed lines at two times the SWH value of each distribution are plotted in the corresponding colors for reference. We see that the combination of positive nonlinearity and non-Markovianity yields extreme events with an intensity that can be tuned with the long-range order parameter, r. For r=7 and positive nonlinearity, the extreme events produced exceed 20 times the SWH.

\begin{figure*}[htbp]
\centering
\includegraphics[width=\textwidth]{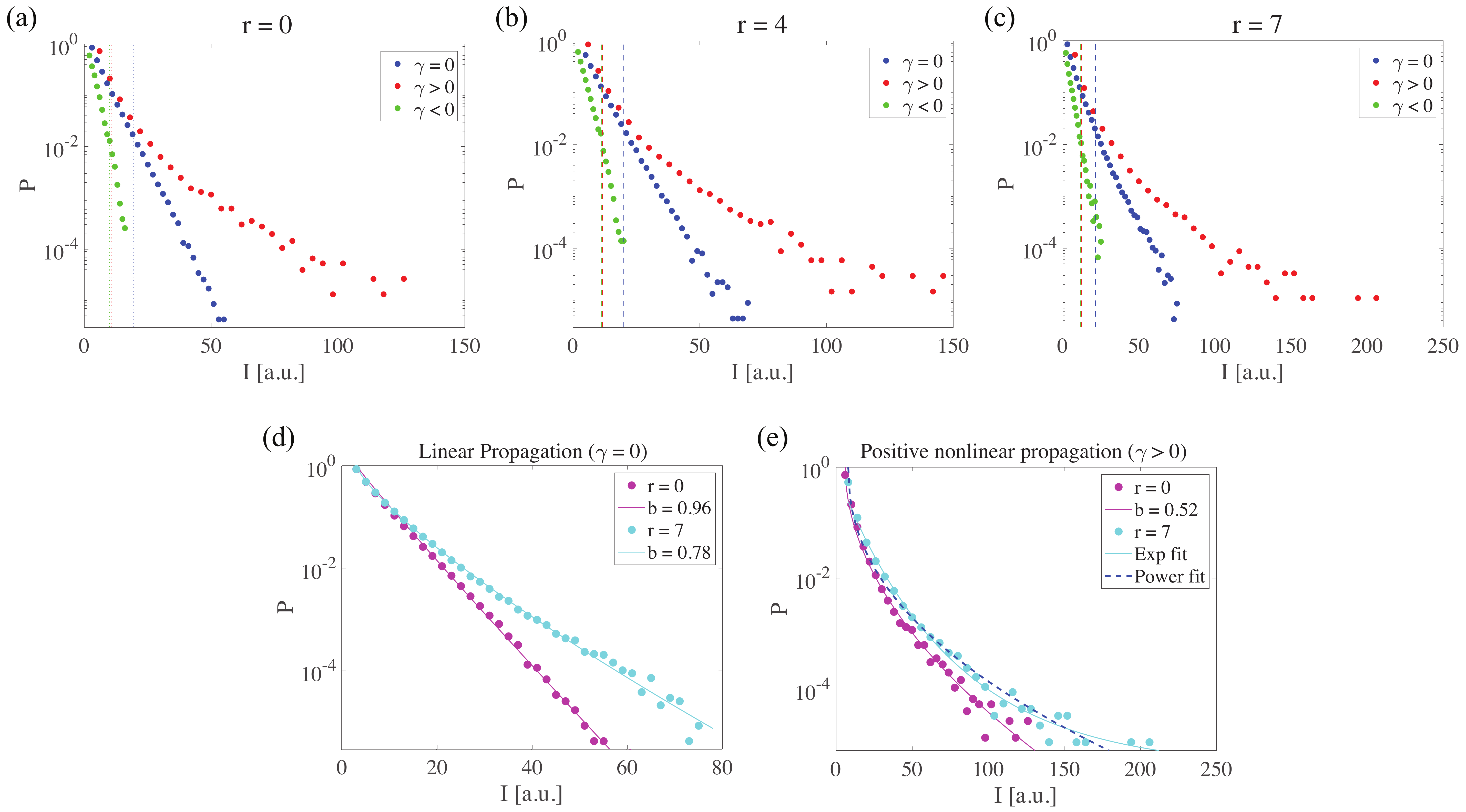}
\caption{\footnotesize\textbf{Experimental results: Intensity probability distributions of measured speckle intensities.} (\textbf{a}) The probability distribution of the intensities generated after propagation through a linear (blue), positive nonlinear (red) and negative nonlinear (green) medium, for a beam imprinted with the r=0 phase mask. Dashed vertical lines at two times the SWH of each distribution, in the corresponding colors, are plotted on top for reference. (\textbf{b}) Similarly, for the r=4 imprinted beam. (\textbf{c}) Similarly, for the r=7 imprinted beam. (\textbf{d}) A comparison of the probability distributions for the r=0 (purple) and r=7 (turquoise) imprinted beams after linear propagation. Both distributions are well fitted by the equation $P_{pow}(I)=\exp(-aI^b+c)$, with $b<1$ for r=7, indicating super-Rayleigh statistics. (\textbf{e}) A comparison of the probability distributions for the r=0 (purple) and r=7 (turquoise) imprinted beams after positive nonlinear propagation. While the r=0 distribution is still well fitted by a power law of the form $P_{pow}$, the r=7 distribution follows a more extreme trend and is well fitted by a function of the form $P_{exp}(I)=\exp[a\exp(-bI)-c]$. The fit to a power law is shown for reference.}
\label{ExpRes}
\end{figure*}

The intensity probability distributions of the experimental results are presented in Fig.~\ref{ExpRes}. Figures~\ref{ExpRes}a-c compare the probability distributions of the light intensity at the crystal output for positive nonlinearity, negative nonlinearity, and no nonlinearity, for varying overlap values. All probability distributions are normalized by the maximum probability of their distribution. Dashed vertical lines indicate the SWH of the distributions. For no overlap (r=0), the linear probability distribution (blue data in Fig.~\ref{ExpRes}a) is approximately a linear line in semilog scale, indicating that it is quite similar to a random, non-correlated field. As expected, when this beam propagates under the influence of positive nonlinearity, the probability distribution develops a long tail, indicating the existence of high intensity, rare events (red data in Fig.~\ref{ExpRes}a) \cite{Bromberg2010,Derevyanko2012a,Sun2012}. This distribution obeys super-Rayleigh statistics, similar to the statistics generated using modulation instability in a positive nonlinear medium. Conversely, when the r=0 imprinted beam propagates under the influence of negative nonlinearity, the probability for high intensity events is suppressed and the probability for low intensity events rises, obeying sub-Rayleigh statistics (green data in Fig.~\ref{ExpRes}a). Similar relations can be seen between the blue, red and green data in figures~\ref{ExpRes}b and c.

However, as the long-range order in the imprinted phase pattern is increased, super-Rayleigh statistics develops even without nonlinear propagation. This can be seen in the blue data in figures~\ref{ExpRes}a-c, and becomes more apparent when comparing the linear intensity probability distribution for the r=0 (purple) and r=7 (turquoise) patterns directly, as in Fig.~\ref{ExpRes}d. The distributions were fitted with a stretched exponential of the form $P_{pow}(I)=\exp(-aI^b+c)$, where I is a vector of intensities, P(I) is the normalized probability to detect a specific intensity, and a, b, and c are parameters. While the logarithm of the linear r=0 distribution fits a power-law with exponent $\simeq1$, the logarithm of the linear r=7 distribution fits a power-law with exponent $0.78$, corresponding to super-Rayleigh statistics.

Though non-Markovianity alone serves to generate extreme events, their intensities are not substantially larger than the intensities observed at the end of the tail of the intensity probability distribution of the r=0 imprinted beam. Yet if a medium with positive nonlinearity is `seeded' with non-Markovian light, the resulting probability distribution has an extremely long tail, longer than that created by either effect on its own. This can be observed in Fig.~\ref{ExpRes}e, which presents a comparison of the distributions obtained from non-Markovian light with no overlap (r=0) and large overlap (r=7) after propagation in a medium with positive nonlinearity. Though both distributions are super-Rayleigh, the distribution obtained for r=7 (turquoise) follows a more extreme trend than that obtained for r=0 (purple). In fact, the distribution obtained for r=7 after nonlinear propagation does not seem to converge to the power-law distribution $P_{pow}$, typical of rogue waves generated via modulation instability in nonlinear media \cite{Bromberg2010,Pierangeli2015,Arecchi2011}. Instead, the distribution is best described by an exponential function of the form $P_{exp}(I)=\exp[a\exp(-bI)-c]$ which resembles the Gumbel distribution \cite{Dean2001}, a particular case of the generalized extreme value (GEV) distributions. Both fits are shown in Fig.~\ref{ExpRes}e.

Overall, the extreme events generated by the combination of non-Markovianity and positive nonlinearity are $\sim2.5$ times more intense than those created just by nonlinearity (best seen when comparing the maximum of the red data in Fig.~\ref{1D}c to the maximum of the red data in Fig.~\ref{1D}a), and $\sim3$ times more intense than those created just by non-Markovianity (red data in Fig.~\ref{1D}c vs. blue data in Fig.~\ref{1D}c). Furthermore, the enhancement of nonlinearity and non-Markovianity combined, $N_{NmNl}$, is about a factor $5$ (red data in Fig.~\ref{1D}c vs. the blue data in Fig.~\ref{1D}a), whereas the two individual enhancements are $N_{Nm}\simeq1.5$ for non-Markovianity alone (blue data in Fig.~\ref{1D}c vs. the blue data in Fig.~\ref{1D}a) and $N_{Nl}\simeq2$ for nonlinearity alone (red data in Fig.~\ref{1D}a vs. the blue data in Fig.~\ref{1D}a). Therefore $N_{NmNl}>N_{Nm}*N_{Nl}$, showing that the two effects are synergistic and their interplay serves to radically increase the wave amplitudes.

\subsection{Simulation results}
\begin{figure*}[htbp]
\centering
\includegraphics[width=\textwidth]{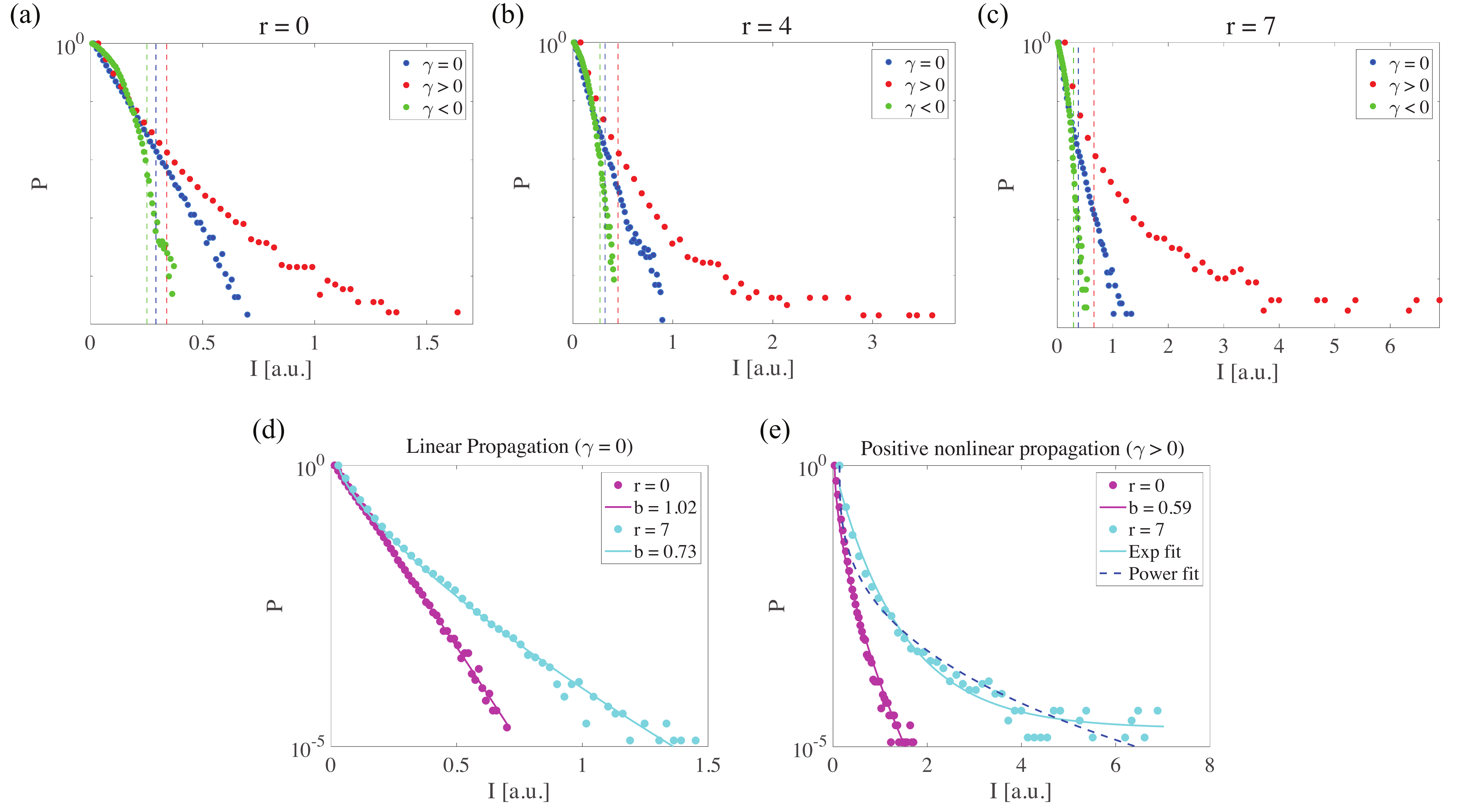}
\caption{\footnotesize\textbf{Simulation results: probability distribution functions of simulated speckle intensities.} (\textbf{a}) The probability distribution of the intensities generated after propagation through a linear (blue), positive nonlinear (red) and negative nonlinear (green) medium, for a beam imprinted with the r=0 phase mask. (\textbf{b}) Similarly, for the r=4 imprinted beam. (\textbf{c}) Similarly, for the r=7 imprinted beam. Dashed vertical lines at two times the SWH of each distribution, in the corresponding colors, are plotted on top for reference. (\textbf{d}) A comparison of the probability distributions for the r=0 (purple) and r=7 (turquoise) imprinted beams after linear propagation. (\textbf{e}) Similarly, for positive nonlinear propagation. All distributions were fitted with the same equations as in Fig.~\ref{ExpRes}. The simulation results show nice qualitative agreement with the experimental results.}
\label{SimPDF}
\end{figure*}

In order to verify the experimental results, we performed numerical simulations of non-Markovian light propagation through nonlinear media. When the nonlinear medium is a photorefractive crystal, the propagation is described by a modified version of the nonlinear Schr\"odinger equation \cite{Delre2009}:
\begin{equation}\label{NLschr}
i\frac{\partial E}{\partial z} + \frac{1}{2k_0n_0}\left[\frac{\partial^2}{\partial x^2} + \frac{\partial^2}{\partial y^2}\right]E - \frac{\gamma E}{1 + |E|^2/I_{bkg}} = 0
\end{equation}
Where $k_0 = \frac{2\pi}{\lambda}$ and $\lambda$ is the wavelength in vacuum, $n_0$ is the linear refractive index of the medium, and $\gamma$ is the voltage dependent nonlinear coefficient of the crystal. $I_{bkg}$ represents the intensity of a second beam incident upon the crystal, of homogeneous incoherent light, which serves as background illumination (see methods). While $I_{bkg}$ was kept constant in the experiment and simulations, the voltage on the crystal was varied such that $\gamma>0$, $\gamma<0$ or $\gamma=0$. To simplify Eq.~\ref{NLschr}, the last term can be approximated as $-\gamma E\left(1 - |E|^2/I_{bkg} + |E|^4/I_{bkg}^2\right)$ for $|E|^2 \ll I_{bkg}$. For positive nonlinearity ($\gamma>0$), this expression includes the well-known Kerr self-focusing contribution as well as a higher-order defocusing contribution which prevents the collapse of small features in $E(x,y)$ to sizes below the resolution of the simulation. For negative nonlinearity ($\gamma<0$), the term $\gamma |E|^2E/I_{bkg}$ causes self-defocusing and the physics is well described with only the third-order term, so the higher order term, $-\gamma|E|^4E/I_{bkg}^2$, was neglected. The intensity distributions after propagation through the medium were analyzed for non-Markovian phase patterns with r=0,4,7 and the corresponding probability distributions are shown in Fig.~\ref{SimPDF}. Again, Fig.~\ref{SimPDF}a-c show a comparison between the distributions resulting from propagation in a linear (blue), positive nonlinear (red) and negative nonlinear (green) medium, for r=0,4,7. Dashed vertical lines in the corresponding colors are plotted at two times the SWH of each distribution. Fig.~\ref{SimPDF}d-e show a comparison between the intensity probability distributions for the r=0 and r=7 imprinted beams, after linear and positive nonlinear propagation, respectively. We see that the simulation results qualitatively agree with the experimental results, showing distributions with increasingly long tails for increasing r values. Moreover, the combination of non-Markovian light and positive nonlinearity results in an extraordinary-long tailed distribution, which is best fit with an exponential of an exponential, as in the experimental results (the fit to a power law is also shown). We note that the effects of nonlinearity on the distributions are stronger in the simulated data than in the experimental data, since we set $\gamma/I_{bkg}$ to a slightly higher value in the simulation in order to make the trends more clear. 

\subsection{1D simulation}
To gain some insight on the reason that non-Markovianity generates long-tailed probability distributions, we take a closer look at the output intensity patterns. In order to make the effect visually clear we use a 1D simulation, which allows for a significantly larger number of statistics compared with a 1D slice of a 2D simulation. The 1D simulation follows the same physics described in section 3.2, yet in one dimension (masks are generated as depicted in Fig.~\ref{PhasePatt}a). The results for a completely random phase mask, as well as for r = 0,4,7,8 after linear (blue), positive nonlinear (red), and negative nonlinear (green) propagation are presented in Fig.~\ref{1Dsim} below. For clarity we have replaced the x-axis, that represented CCD pixles in our previous plots, with the frequencies (denoted as $k$) that represent the Fourier-conjugate variable of the pixels of the SLM. Fig.~\ref{1Dsim}a presents the case of a completely random phase mask, where the numbers 1-9 are randomly assigned to pixels on the SLM. The far-field pattern is a Rayleigh-distributed speckle pattern for linear propagation, super-Rayleigh distributed speckle pattern for positive nonlinear propagation, and sub-Rayleigh distributed speckles for negative nonlinear propagation, as has been demonstrated previously \cite{Bromberg2010,Frostig2017} (for plots of the corresponding probability distributions see Fig.~S1).

As in the 2D case, the r=0 case represents the non-Markovian phase mask with the least amount of order, since the phase of a given pixel is statistically dependent only on the other pixels within its own 1x9 puzzle. While the far-field is still close to a random speckle pattern (see Fig.~\ref{1Dsim}b), we can gain some understanding of its nature by considering how often numbers can repeat in the near-field phase mask. The constraint that the numbers 1-9 appear once in each puzzle suppresses the probability of very low frequencies, since numbers must repeat at least every 17 entries. This manifests in a dip in the probability of high intensity speckles near $k=0$ (the fall off is slow, rather than sharp, due to the random pattern superposed). This attenuation drives some energy into the rest of the spectrum, causing some higher intensity speckles to appear and slightly elongating the probability distributions. Nevertheless the distributions do not significantly deviate from those of a completely random phase mask (see Fig.~S1a and.~S1b).

As the value of r increases, more long-range order appears in the applied phase masks, with r=4 being an intermediate case (Fig.~\ref{1Dsim}c). The dip in the low frequencies widens, pushing more energy into the speckles in the intermediate frequency region. While the effect is less pronounced for the linear case, it is significant for the positive nonlinear case. As a result, the probability distributions, in particular the positive nonlinear distribution, become appreciably elongated.

The most interesting case is r=7, shown in Fig.~\ref{1Dsim}d. This mask is the non-Markovian phase mask with the most order, yet random assignment is maintained. Since numbers must repeat every 8-10 entries, the most likely frequencies are $k=1/9$ and adjacent frequencies, as well as the second and third harmonic of these frequencies. As a result, the output pattern has three regions in which the likelihood of high-intensity speckles is higher, centered around $k=1/9$, $k=2/9$ and $k=3/9$. The region centered at the fundamental frequency, $k=1/9$, is the most narrow and sharpest, while the harmonic regions become gradually wider and less sharp. In between the high-probability regions there are dips in the pattern, or regions of low-probability of high-intensity speckles. This pattern of low- and high-probability regions is most pronounced for the positive nonlinearity case, where it contributes to the emergence of an extraordinarily long probability distribution (see Fig.~S1d), but is present also for the linear and negative nonlinear cases.

Finally, the r=8 case represents a completely ordered phase mask. Order emerges because when assigning the ninth pixel of a new puzzle whose previous eight pixels overlap with an existing puzzle, only one `free' number out of the numbers 1-9 is left. Therefore, the first puzzle is randomly assigned and it is repeated periodically thereafter until the whole phase mask is assigned. The near-field pattern can thus be described as a convolution of a Dirac comb and a random phase pattern of a length of 9 entries. As a result, the far-field pattern after linear propagation, shown in Fig.~\ref{1Dsim}e, is a Dirac comb multiplied by a random intensity pattern, resulting in a comb with random teeth height. The width of the teeth is then broadened by diffraction during linear propagation (the patterns after nonlinear propagation are not presented for this case for clarity). We note that for r=8 the probability distribution is again a Rayleigh distribution (see Fig.~S1e). Therefore the more long-range order exists in the non-Markovian mask, the more elongated the intensity distribution. Yet if the random assignment disappears completely, so does the long-tailed intensity distribution.

\begin{figure}[htbp]
\centering
\includegraphics[width=\linewidth]{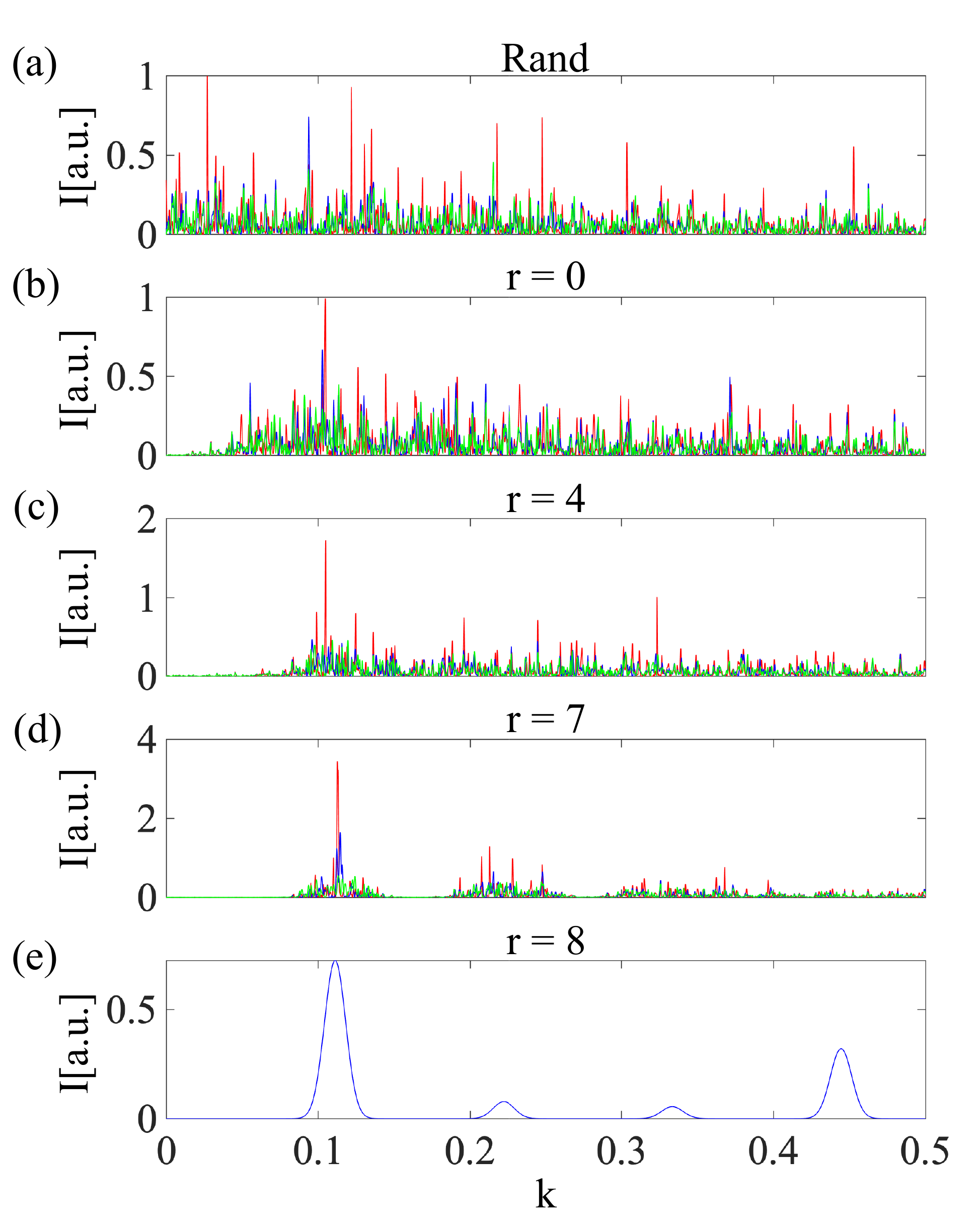}
\caption{\footnotesize\textbf{1D simulation results: The output intensity patterns for different 1D phase masks, after propagation through a linear (blue), positive nonlinear (red) and negative nonlinear (green) medium.} (\textbf{a}) For a completely random phase mask. (\textbf{b}) For the r=0 phase mask. (\textbf{c}) For the r=4 phase mask. (\textbf{d}) For the r=7 phase mask. (\textbf{a}) For the r=8 phase mask. As order is increased, regions of high probability for high intensity events emerge, causing the elongation of the intensity distribution (see Fig.~S1). Nevertheless when random assignment disappears completely, as in the r=8 phase mask, the Rayleigh intensity distribution reemerges. The nonlinear cases are not shown for r=8 for clarity.
\label{1Dsim}}
\end{figure}

\section{Discussion}
In this work we have observed, for the first time, the generation of extreme, rare events from a non-Markovian light source. Moreover, we have shown that when a nonlinear medium is seeded with non-Markovian light, the interplay between these two effects leads to the generation of exceptionally large rogue events. The resulting intensity distribution follows a more extreme trend than distributions conventionally generated in rogue wave experiments, such as those obtained with nonlinearity alone.

The unique statistical properties of our system, as well as its tunability and experimental simplicity, make it a convenient test-bed for studying the physics of extreme events. Furthermore, as the probability distributions created by our system resemble generalized extreme value distributions, it may be a suitable model for extreme-value systems that could not be modeled previously.

\section{Materials and methods}
\subsection{Experimental implementation}
Light from a 532nm CW laser (Coherent Verdi) was magnified with a 4x telescope, and reflected off a two-dimensional liquid-crystal spatial-light modulator (Hamamatsu LCOS-SLM x10468). The SLM was used to imprint the beam with a non-Markovian phase mask. The outgoing field was Fourier-transformed using a 500mm focal length lens in order to obtain a non-Markovian intensity pattern (fully-developed speckles), and inserted into a photorefractive crystal (SBN:60) with a 5mm x 5mm cross-section and 10mm length. The output facet of the crystal was imaged onto a CCD camera (UI-1250 IDS) and the recorded intensity patterns were analyzed. The background illumination for the crystal (represented by $I_{bkg}$ in Eq.~\ref{NLschr}) was generated by splitting the main beam before the telescope, and passing it through a rotating diffuser in order to destroy its coherence. The intensity of the background illumination was $\sim$4 times the intensity of the non-Markovian light. The SLM pixels were binned into 5-by-5 macro pixels. Each phase mask was displayed on the SLM and the intensity probability distribution was computed from the histogram of the pixel values in the corresponding CCD image, with 50 equally-spaced bins. In distributions where there were less than 50 discrete pixel values (such as the negative nonlinearity case for low r values), the maximum possible number of bins was used. Each distribution was normalized to its maximum value. Pixels in the image with a value of zero were eliminated in order to use a log plot. For each r value, different realizations of the non-Markovian phase mask were generated by computing different solutions of the Sudoku puzzles. The measurements with linear propagation were performed without applying voltage to the crystal, and were averaged over 20 realizations of the non-Markovian phase mask for each r value. The voltage applied for positive nonlinearity measurements was $\sim$0.7kV, and the results were averaged over four realizations, with 50 measurements for each realization. The voltage applied for negative nonlinearity measurements was -0.5kV, and the results were averaged over two realizations. In all three cases, the number of realizations was chosen to maintain reasonable experimental run times after verifying numerically and experimentally that the amount of statistics produced created smooth intensity distributions.  

\subsection{2D simulation implementation}
The non-Markovian phase masks used were either the same as those used in the experiment, or generated in the same way. Nonlinear propagation was simulated using the split-step method. Time-domain was not addressed specifically in the simulation, as this would result in impractical run times. 
Defocusing nonlinearity was modeled as Kerr-type, with the nonlinear operator $\hat{N}=-\gamma/I_{bkg}|E|^2$. Focusing nonlinearity was modeled with a higher-order defocusing term, $\hat{N}=\gamma/I_{bkg}(|E|^2-1/I_{bkg}|E|^4)$. The higher order term introduces saturation of the self-focusing process, as occurs in practice in photorefractive crystals \cite{Bian1997}, and prevents the collapse of the speckles to sizes below the resolution of the simulation. The results were averaged over 20 realizations for linear, positive nonlinear, and negative nonlinear propagation. All histograms were computed using 50 equally spaced bins and were normalized to their maximum values. Zeros values were removed due the log plot.

\subsection{1D simulation implementation}
The simulation implementation followed that of the 2D simulation, but in a single spatial dimension ((1+1)D geometry). The results were averaged over 100 realizations for the completely random case and for r=0,4,7. For r=8 the results were averaged over 1000 realizations since the r=8 output pattern contains significantly less statistics than other r values.  All histograms were computed using 25 equally spaced bins and were normalized to their maximum values. Zeros values were removed due the log plot.

\bigskip
\bigskip
\noindent\textbf{\large Acknowledgements}

\noindent The authors dedicate this paper to the late Prof. Yaron Silberberg, a great scientist, mentor, and person. The authors thank O. Raz, I. Kantor, D. Mukamel, M. Segev and Y. Gilead for helpful discussions. This work was supported by grants from the CAPES/Weizmann cooperation program, CNPq (Universal Grant No. 483983/2013-6), the Icore program of the ISF, and the Crown Photonics Center.

\bigskip
\noindent\textbf{\large Author contributions}

\noindent H.F. and I.V. contributed equally to this work. I.V. conceived the idea. H.F., I.V. and Y.S. designed the experiment, analyzed the data and prepared the manuscript. H.F. and I.V. performed the experiment. H.F., I.V. and R.F. worked on simulations. The crystal and technical support was provided by H.H.S.

\bigskip
\noindent\textbf{\large Competing financial interest}

\noindent The author declare no competing financial interests.


\bibliographystyle{ieeetr}
\bibliography{NLsudoku}

\begin{thebibliography}{10}

\bibitem{Dysthe2008}
K.~Dysthe, H.~E. Krogstad, and P.~M{\"u}ller, ``Oceanic {{Rogue Waves}},'' {\em
  Annu. Rev. Fluid Mech.}, vol.~40, no.~1, pp.~287--310, 2008.

\bibitem{Solli2007}
D.~R. Solli, C.~Ropers, P.~Koonath, and B.~Jalali, ``Optical rogue waves,''
  {\em Nature}, vol.~450, pp.~1054--1057, Dec. 2007.

\bibitem{Baronio2012}
F.~Baronio, A.~Degasperis, M.~Conforti, and S.~Wabnitz, ``Solutions of the
  {{Vector Nonlinear Schr{\"o}dinger Equations}}: {{Evidence}} for
  {{Deterministic Rogue Waves}},'' {\em Phys. Rev. Lett.}, vol.~109, July 2012.

\bibitem{Toenger2015}
S.~Toenger, T.~Godin, C.~Billet, F.~Dias, M.~Erkintalo, G.~Genty, and J.~M.
  Dudley, ``Emergent rogue wave structures and statistics in spontaneous
  modulation instability,'' {\em Sci. Rep.}, vol.~5, p.~10380, May 2015.

\bibitem{Kibler2010}
B.~Kibler, J.~Fatome, C.~Finot, G.~Millot, F.~Dias, G.~Genty, N.~Akhmediev, and
  J.~M. Dudley, ``The {{Peregrine}} soliton in nonlinear fibre optics,'' {\em
  Nat. Phys.}, vol.~6, pp.~790--795, Oct. 2010.

\bibitem{Mussot2009}
A.~Mussot, A.~Kudlinski, M.~Kolobov, E.~Louvergneaux, M.~Douay, and M.~Taki,
  ``Observation of extreme temporal events in cw-pumped supercontinuum,'' {\em
  Opt. Express}, vol.~17, pp.~17010--17015, Sep 2009.

\bibitem{Erkintalo2009}
M.~Erkintalo, G.~Genty, and J.~M. Dudley, ``Rogue-wave-like characteristics in
  femtosecond supercontinuum generation,'' {\em Opt. Lett.}, vol.~34,
  pp.~2468--2470, Aug 2009.

\bibitem{Hammani2008}
K.~Hammani, C.~Finot, J.~M. Dudley, and G.~Millot, ``Optical rogue-wave-like
  extreme value fluctuations in fiber raman amplifiers,'' {\em Opt. Express},
  vol.~16, pp.~16467--16474, Oct 2008.

\bibitem{Pierangeli2015}
D.~Pierangeli, F.~Di~Mei, C.~Conti, J.~Agranat, A.\, and E.~DelRe, ``Spatial
  {{Rogue Waves}} in {{Photorefractive Ferroelectrics}},'' {\em Phys. Rev.
  Lett.}, vol.~115, Aug. 2015.

\bibitem{Montina2009}
A.~Montina, U.~Bortolozzo, S.~Residori, and F.~T. Arecchi, ``Non-{{Gaussian
  Statistics}} and {{Extreme Waves}} in a {{Nonlinear Optical Cavity}},'' {\em
  Phys. Rev. Lett.}, vol.~103, Oct. 2009.

\bibitem{Dubietis2011}
D.~Majus, V.~Jukna, G.~Valiulis, D.~Faccio, and A.~Dubietis, ``Spatiotemporal
  rogue events in femtosecond filamentation,'' {\em Phys. Rev. A}, vol.~83,
  p.~025802, Feb 2011.

\bibitem{Birkholz2013}
S.~Birkholz, E.~T.~J. Nibbering, C.~Br\'ee, S.~Skupin, A.~Demircan, G.~Genty,
  and G.~Steinmeyer, ``Spatiotemporal rogue events in optical multiple
  filamentation,'' {\em Phys. Rev. Lett.}, vol.~111, p.~243903, Dec 2013.

\bibitem{Agafontsev2015}
D.~S. Agafontsev and V.~E. Zakharov, ``Integrable turbulence and formation of
  rogue waves,'' {\em Nonlinearity}, vol.~28, pp.~2791--2821, jul 2015.

\bibitem{Horowitz1997}
M.~Horowitz, Y.~Barad, and Y.~Silberberg, ``Noiselike pulses with a broadband
  spectrum generated from an erbium-doped fiber laser,'' {\em Opt. Lett.},
  vol.~22, pp.~799--801, Jun 1997.

\bibitem{suret2016}
P.~Suret, R.~El~Koussaifi, A.~Tikan, C.~Evain, S.~Randoux, C.~Szwaj, and
  S.~Bielawski, ``Single-shot observation of optical rogue waves in integrable
  turbulence using time microscopy,'' {\em Nature communications}, vol.~7,
  p.~13136, 2016.

\bibitem{Coulibaly2019}
S.~Coulibaly, M.~Taki, A.~Bendahmane, G.~Millot, B.~Kibler, and M.~G. Clerc,
  ``Turbulence-induced rogue waves in kerr resonators,'' {\em Phys. Rev. X},
  vol.~9, p.~011054, Mar 2019.

\bibitem{Lecaplain2014}
C.~Lecaplain and P.~Grelu, ``Rogue waves among noiselike-pulse laser emission:
  An experimental investigation,'' {\em Phys. Rev. A}, vol.~90, p.~013805, Jul
  2014.

\bibitem{Lecaplain2012}
C.~Lecaplain, P.~Grelu, J.~M. Soto-Crespo, and N.~Akhmediev, ``Dissipative
  rogue waves generated by chaotic pulse bunching in a mode-locked laser,''
  {\em Phys. Rev. Lett.}, vol.~108, p.~233901, Jun 2012.

\bibitem{SotoCrespo2011}
J.~M. Soto-Crespo, P.~Grelu, and N.~Akhmediev, ``Dissipative rogue waves:
  Extreme pulses generated by passively mode-locked lasers,'' {\em Phys. Rev.
  E}, vol.~84, p.~016604, Jul 2011.

\bibitem{Liu2016}
M.~Liu, A.-P. Luo, W.-C. Xu, and Z.-C. Luo, ``Dissipative rogue waves induced
  by soliton explosions in an ultrafast fiber laser,'' {\em Opt. Lett.},
  vol.~41, pp.~3912--3915, Sep 2016.

\bibitem{onorato2013}
M.~Onorato, S.~Residori, U.~Bortolozzo, A.~Montina, and F.~Arecchi, ``Rogue
  waves and their generating mechanisms in different physical contexts,'' {\em
  Physics Reports}, vol.~528, no.~2, pp.~47--89, 2013.

\bibitem{Dudley2014}
J.~M. Dudley, F.~Dias, M.~Erkintalo, and G.~Genty, ``Instabilities, breathers
  and rogue waves in optics,'' {\em Nat. Photon.}, vol.~8, pp.~755--764, Sept.
  2014.

\bibitem{Dudley2019}
J.~M. Dudley, G.~Genty, A.~Mussot, A.~Chabchoub, and F.~Dias, ``Rogue waves and
  analogies in optics and oceanography,'' {\em Nature Reviews Physics},
  pp.~1--15, 2019.

\bibitem{Weng1992}
L.~Weng, J.~M. Reid, P.~M. Shankar, K.~Soetanto, and X.~M. Lu, ``Nonuniform
  phase distribution in ultrasound speckle analysis. {{I}}. {{Background}} and
  experimental demonstration,'' {\em IEEE Transactions on Ultrasonics,
  Ferroelectrics, and Frequency Control}, vol.~39, pp.~352--359, May 1992.

\bibitem{goodman2000}
J.~W. Goodman, {\em Statistical {{Optics}}}.
\newblock New York: {Wiley-Interscience}, 1 edition~ed., Aug. 2000.

\bibitem{Nieuwenhuizen1995}
T.~M. Nieuwenhuizen and M.~C.~W. Van~Rossum, ``Intensity distributions of waves
  transmitted through a multiple scattering medium,'' {\em Phys. Rev. E},
  vol.~74, no.~14, p.~2674, 1995.

\bibitem{Metzger2014}
J.~J. Metzger, R.~Fleischmann, and T.~Geisel, ``Statistics of {{Extreme Waves}}
  in {{Random Media}},'' {\em Phys. Rev. Lett.}, vol.~112, May 2014.

\bibitem{Hohmann2010}
R.~H{\"o}hmann, U.~Kuhl, H.-J. St{\"o}ckmann, L.~Kaplan, and E.~J. Heller,
  ``Freak {{Waves}} in the {{Linear Regime}}: {{A Microwave Study}},'' {\em
  Phys. Rev. Lett.}, vol.~104, Mar. 2010.

\bibitem{Strudley2013}
T.~Strudley, T.~Zehender, C.~Blejean, E.~P. A.~M. Bakkers, and O.~L. Muskens,
  ``Mesoscopic light transport by very strong collective multiple scattering in
  nanowire mats,'' {\em Nat. Photon.}, vol.~7, pp.~413--418, Apr. 2013.

\bibitem{Mathis2015}
A.~Mathis, L.~Froehly, S.~Toenger, F.~Dias, G.~Genty, and J.~M. Dudley,
  ``Caustics and {{Rogue Waves}} in an {{Optical Sea}},'' {\em Sci. Rep.},
  vol.~5, p.~12822, Aug. 2015.

\bibitem{Arecchi2011}
F.~T. Arecchi, U.~Bortolozzo, A.~Montina, and S.~Residori, ``Granularity and
  {{Inhomogeneity Are}} the {{Joint Generators}} of {{Optical Rogue Waves}},''
  {\em Phys. Rev. Lett.}, vol.~106, Apr. 2011.

\bibitem{Apostol2003}
A.~Apostol and A.~Dogariu, ``Spatial {{Correlations}} in the {{Near Field}} of
  {{Random Media}},'' {\em Phys. Rev. Lett.}, vol.~91, p.~093901, Aug. 2003.

\bibitem{Safari2017}
A.~Safari, R.~Fickler, M.~J. Padgett, and R.~W. Boyd, ``Generation of caustics
  and rogue waves from nonlinear instability,'' {\em Phys. Rev. Lett.},
  vol.~119, p.~203901, Nov 2017.

\bibitem{Liu2015}
C.~Liu, R.~E.~C. {van der Wel}, N.~Rotenberg, L.~Kuipers, T.~F. Krauss,
  A.~Di~Falco, and A.~Fratalocchi, ``Triggering extreme events at the nanoscale
  in photonic seas,'' {\em Nat. Phys.}, vol.~11, pp.~358--363, Mar. 2015.

\bibitem{Bromberg2014}
Y.~Bromberg and H.~Cao, ``Generating {{Non}}-{{Rayleigh Speckles}} with
  {{Tailored Intensity Statistics}},'' {\em Phys. Rev. Lett.}, vol.~112, May
  2014.

\bibitem{Fischer2015}
R.~Fischer, I.~Vidal, D.~Gilboa, B.~Correia, Ricardo~R.\, A.~C.
  Ribeiro-Teixeira, S.~D. Prado, J.~Hickman, and Y.~Silberberg, ``Light with
  {{Tunable Non}}-{{Markovian Phase Imprint}},'' {\em Phys. Rev. Lett.},
  vol.~115, p.~073901, Aug. 2015.

\bibitem{Eichelkraut2015}
T.~Eichelkraut and A.~Szameit, ``Photonics: {{Random}} sudoku light,'' {\em
  Nature}, vol.~526, pp.~643--644, Oct. 2015.

\bibitem{delsole2000}
T.~DelSole, ``A fundamental limitation of markov models,'' {\em Journal of the
  atmospheric sciences}, vol.~57, no.~13, pp.~2158--2168, 2000.

\bibitem{sura2003}
P.~Sura, ``Stochastic analysis of southern and pacific ocean sea surface
  winds,'' {\em Journal of the atmospheric sciences}, vol.~60, no.~4,
  pp.~654--666, 2003.

\bibitem{thompson2014}
W.~F. Thompson, A.~H. Monahan, and D.~Crommelin, ``Parametric estimation of the
  stochastic dynamics of sea surface winds,'' {\em Journal of the Atmospheric
  Sciences}, vol.~71, no.~9, pp.~3465--3483, 2014.

\bibitem{keen2015}
D.~A. Keen and A.~L. Goodwin, ``The crystallography of correlated disorder,''
  {\em Nature}, vol.~521, no.~7552, p.~303, 2015.

\bibitem{Schorr2003}
R.~Schorr and H.~Rieger, ``Universal properties of shortest paths in
  isotropically correlated random potentials,'' {\em The European Physical
  Journal B-Condensed Matter and Complex Systems}, vol.~33, no.~3,
  pp.~347--354, 2003.

\bibitem{akhmediev2016}
N.~Akhmediev, B.~Kibler, F.~Baronio, M.~Beli{\'c}, W.-P. Zhong, Y.~Zhang,
  W.~Chang, J.~M. Soto-Crespo, P.~Vouzas, P.~Grelu, {\em et~al.}, ``Roadmap on
  optical rogue waves and extreme events,'' {\em Journal of Optics}, vol.~18,
  no.~6, p.~063001, 2016.

\bibitem{Bromberg2010}
Y.~Bromberg, Y.~Lahini, E.~Small, and Y.~Silberberg, ``Hanbury {{Brown}} and
  {{Twiss}} interferometry with interacting photons,'' {\em Nat. Photonics},
  vol.~4, pp.~721--726, 2010.

\bibitem{Derevyanko2012a}
S.~Derevyanko and E.~Small, ``Nonlinear propagation of an optical speckle
  field,'' {\em Phys. Rev. A}, vol.~053816, no.~85, pp.~1--9, 2012.

\bibitem{Sun2012}
C.~Sun, S.~Jia, C.~Barsi, S.~Rica, A.~Picozzi, and J.~W. Fleischer,
  ``Observation of the kinetic condensation of classical waves,'' {\em Nat.
  Phys.}, vol.~8, pp.~470--474, Apr. 2012.

\bibitem{Dean2001}
D.~S. Dean and S.~N. Majumdar, ``Extreme-value statistics of hierarchically
  correlated variables deviation from {{Gumbel}} statistics and anomalous
  persistence,'' {\em Phy. Rev. E}, vol.~64, Sept. 2001.

\bibitem{Delre2009}
E.~DelRe, B.~Crosignani, and P.~Di~Porto, ``Chapter 3 {{Photorefractive
  Solitons}} and {{Their Underlying Nonlocal Physics}},'' in {\em Progress in
  {{Optics}}}, vol.~53, pp.~153--200, {Elsevier}, 2009.

\bibitem{Frostig2017}
H.~Frostig, E.~Small, A.~Daniel, P.~Oulevey, S.~Derevyanko, and Y.~Silberberg,
  ``Focusing light by wavefront shaping through disorder and nonlinearity,''
  {\em Optica}, vol.~4, no.~9, pp.~1073--1079, 2017.

\bibitem{Bian1997}
S.~Bian, J.~Frejlich, and K.~H. Ringhofer, ``Photorefractive saturable
  kerr-type nonlinearity in photovoltaic crystals,'' {\em Phys. Rev. Lett.},
  vol.~78, pp.~4035--4038, May 1997.

\end{thebibliography}


\begin{thebibliography}{1}

\bibitem{Akhmediev2016}
N.~Akhmediev, B.~Kibler, F.~Baronio, M.~Beli{\'c}, W.-P. Zhong, Y.~Zhang,
  W.~Chang, J.~M. Soto-Crespo, P.~Vouzas, P.~Grelu, {\em et~al.}, ``Roadmap on
  optical rogue waves and extreme events,'' {\em Journal of Optics}, vol.~18,
  no.~6, p.~063001, 2016.

\bibitem{onorato2013}
M.~Onorato, S.~Residori, U.~Bortolozzo, A.~Montina, and F.~Arecchi, ``Rogue
  waves and their generating mechanisms in different physical contexts,'' {\em
  Physics Reports}, vol.~528, no.~2, pp.~47--89, 2013.

\bibitem{Dudley2014}
J.~M. Dudley, F.~Dias, M.~Erkintalo, and G.~Genty, ``Instabilities, breathers
  and rogue waves in optics,'' {\em Nat. Photon.}, vol.~8, pp.~755--764, Sept.
  2014.

\end{thebibliography}

\end{document}


\title{Observation of rogue events in non-Markovian light: supplementary material}

\author{Hadas Frostig$^{1,\dag,*}$, Itamar Vidal$^{1,2,\dag}$,Robert Fischer$^3$, Hanan Herzig Sheinfux$^4$, and Yaron Silberberg$^{1,5}$}

\affil{\footnotesize $^1$Department of Physics of complex Systems, Weizmann Institute of Science, Rehovot 76100, Israel \\$^2$ 2LOQEF,  57039-739 Maceió, Alagoas, Brazil \\$^3$CPGEI, Federal University of Technology - Paran\'a, 80230-901 Curitiba, PR, Brazil \\$^4$ Physics Department, Technion, Israel Institute of Technology, Haifa 32000, Israel \\$^5$ Deceased}

\twocolumn[
  \maketitle
\begin{onecolabstract}
This document provides supplementary information to ``Observation of rogue events in non-Markovian light''. The supplementary material includes two sections. The first section presents the intensity probability distributions of the one-dimensional simulated data presented in Fig.~6 of the main text. The second section presents the propagation traces of the one-dimensional simulated data through the linear and nonlinear media.
\vspace{1cm}
\end{onecolabstract}
]

\section{Intensity probability distributions of the one-dimensional simulated data}
Below we present the intensity probability distributions computed from the 1D simulation, corresponding to Fig.~6 of the main text. Figures~S1a-e show the distributions generated by the totally random phase mask, the r=0 mask, the r=4 mask, the r=7 mask, and the r=8 mask, respectively. Each sub figure displays the distribution after propagation through a linear (blue), positive nonlinear (red) and negative nonlinear (green) medium. As in the 2D simulation, the distribution becomes more elongated as r increases, and the combination of positive nonlinear propagation and large r values serves to create extremely long-tailed distributions. Yet when the order is increased further and the phase mask becomes completely ordered, as occurs for r=8 (Fig~S1e), the distribution becomes linear just like in the completely random case (Fig~S1a).
\blfootnote{\dag These authors contributed equally to this work.} 
\blfootnote{*Corresponding author: hfrostig@bu.edu.}

\begin{figure}
\includegraphics[width=\columnwidth]{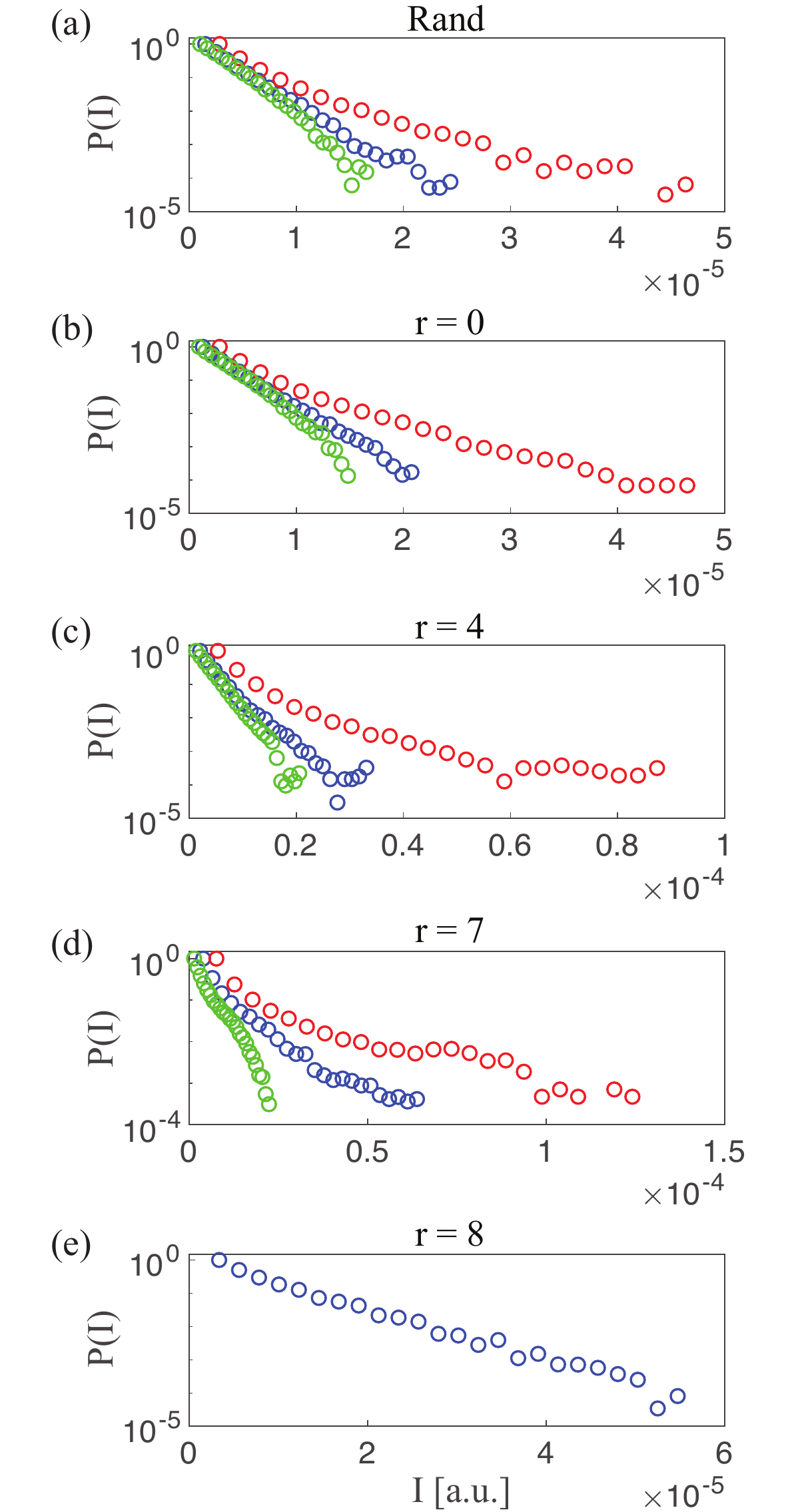}
\caption{\footnotesize\textbf{1D simulation results: The intensity probability distributions for different 1D phase masks, after propagation through a linear (blue), positive nonlinear (red) and negative nonlinear (green) medium.} (\textbf{a}) For a completely random phase mask. (\textbf{b}) For the r=0 phase mask. (\textbf{c}) For the r=4 phase mask. (\textbf{d}) For the r=7 phase mask. (\textbf{e}) For the r=8 phase mask. The trends follow those of the 2D simulation, showing more elongated distributions for larger r values. Yet when the mask becomes completely ordered, for r=8, the distribution becomes linear again, as in the completely random case.
\label{1D_PDFs}}	
\end{figure}

\section{Propagation traces of the one-dimensional simulated data}
Below we show examples of simulated propagation traces of the one-dimensional far-field pattern generated by the r=7 phase mask. Propagation traces through negative nonlinear (Fig.~S2a), linear (Fig.~S2b) and positive nonlinear (Fig.~S2c) media are presented. z=0 represents the input facet of the photorefractive crystal and z=1 represents the output facet. Several examples of waves whose intensity exceeds two times the SWH are marked with white arrows. Intensities were normalized to the maximum intensity in the pattern at position z=0. We see that the rogue events generated in all three types of media are transient. The traces in Fig.~S2a and S2c follow the typical trends of the nonlinear Schr\"odinger equation \cite{Akhmediev2016,onorato2013,Dudley2014}.

\begin{figure*}
\includegraphics[width=\textwidth]{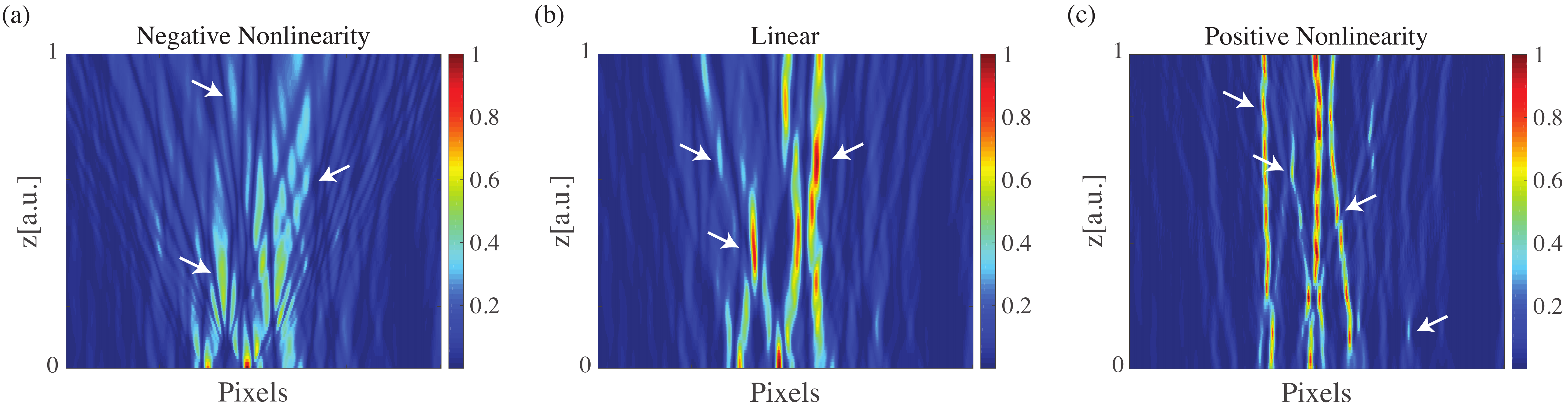}
\caption{\footnotesize\textbf{1D simulation results: Propagation traces of the far-field patterns of the r=7 non-Markovian phase mask.} The propagation is shown from the input facet of the photorefractive crystal (r=0) to the output facet (r=1). (\textbf{a}) Propagation through a negative nonlinear medium. (\textbf{b}) Propagation through a linear medium. (\textbf{c}) Propagation through a positive nonlinear medium. Examples of waves exceeding two times the SWH are marked with white arrows. We note that the generated rogue events for all three mediums are transient.
\label{PropMats}}	
\end{figure*}

\bibliographystyle{ieeetr}
\bibliography{NLsudoku}